\documentclass[useAMS,usenatbib]{mn2e}
\usepackage[T1]{fontenc}
\usepackage{aecompl}
\usepackage{txfonts}
\usepackage{natbib}
\usepackage{aas_macros}
\usepackage[all]{xy}

 \usepackage{ifthen}
 \ifx\pdftexversion\undefined
 \usepackage[dvips]{graphicx}
 \else
 \usepackage[pdftex]{graphicx}
 \usepackage{epstopdf}
 \usepackage{thumbpdf}
 \fi

\usepackage{hyperref}
\usepackage[usenames,dvipsnames]{color}
\definecolor{darkblue}{rgb}{0,0,.5}
\hypersetup{bookmarksopen,
			pdfstartview={FitH},
			colorlinks=true,
			breaklinks=true,
			linkcolor=darkblue,
			menucolor=darkblue,
			urlcolor=Red,
			citecolor=Green,
			linkcolor=blue}
\usepackage[all]{hypcap}

\def\del#1{{}}
\newcommand{\dd}{\mathrm{d}}
\newcommand{\eqref}[1]{(\ref{#1})}

\title[Extreme value statistics of CMB lensing]{Extreme value statistics of CMB lensing deflection angles}
\author[Philipp M. Merkel and Bj{\"o}rn Malte Sch\"afer]
{Philipp M. Merkel$^1$\thanks{e-mail: philipp.merkel@uni-heidelberg.de} and Bj{\"o}rn Malte Sch\"afer$^2$\\
${}^1$Institut f{\"u}r Theoretische Astrophysik, Zentrum f{\"u}r Astronomie, Universit{\"a}t Heidelberg, Philosophenweg 12, 69120 Heidelberg, Germany\\
${}^2$Astronomisches Recheninstitut, Zentrum f{\"u}r Astronomie, Universit{\"a}t Heidelberg, Philosophenweg 12, 69120 Heidelberg, Germany}

\begin{document}
% \onecolumn
\pagerange{\pageref{firstpage}--\pageref{lastpage}}
\pubyear{2014}
\maketitle
\label{firstpage}

% --- abstract --- %
\begin{abstract}
The smaller the angular scales on which the anisotropies of the cosmic microwave background (CMB) are probed the more important their distortion due to gravitational lensing becomes.
Here we investigate the maxima and minima of the CMB lensing deflection field using general extreme value statistics. Since general extreme value statistics applies to uncorrelated data in first place we consider appropriately low-pass filtered deflection maps. Besides the suppression of correlations filtering is required for another reason: The lensing field itself is not directly observable but needs to be (statistically) reconstructed from the lensed CMB by means of a quadratic estimator. This reconstruction, though, is noise dominated and therefore requires smoothing, too. In idealized Gaussian realizations as well as in realistically reconstructed data we find that both maxima and minima of the deflection angle components follow consistently a general extreme value distribution of Weibull-type. However, its shape, location and scale parameters vary significantly between different realizations. The statistics' potential power to constrain cosmological models appears therefore rather limited. 
\end{abstract}

% --- keywords --- %
\begin{keywords}
cosmic background radiation -- gravitational lensing: weak -- methods: numerical
\end{keywords}

% --- introduction --- %
\section{Introduction}
\label{sec_introduction}

While traversing the Universe photons of the Cosmic Microwave Background (CMB) are deflected by the intervening large scale structure: They are gravitationally lensed.
Lensing widens the size distribution of hot and cold spots imprinted at the surface of last scattering.  At the primary acoustic peaks this effect amounts to several per cent and dominates in the damping tail transferring power from larger to smaller angular scales. CMB lensing has been measured in a number of experiments, most recently in the \textit{Planck} data \citep{2014A&A...571A..17P}.

Precise knowledge of the lensed CMB temperature and polarization power spectra allows to lift several parameter degeneracies \citep[eg.][]{2006PhRvD..74l3002S}. In particular the projection degeneracy of the primary CMB between spatial curvature and a cosmological constant can be alleviated through lensing \citep{2006ApJ...650L..13H}. Moreover, CMB lensing is expected to contain important information about the onset of late-time cosmic acceleration \citep{2006PhRvD..74j3510A} and has recently been reconsidered in the investigation of early dark energy \citep{2013PhRvD..87h3009P}.
Regarding the statistics of the observed, i.e. lensed, CMB anisotropies there are two important differences with respect to the primordial fluctuations. Firstly, the lensed CMB is necessarily non-Gaussian, even in case of Gaussian temperature fluctuations, because lensing is a nonlinear effect, and secondly it is not statistically isotropic. The off-diagonal elements of its covariance can be used to reconstruct (in a statistical sense) the lensing deflection field \citep{2001ApJ...557L..79H,2003PhRvD..67h3002O,2003PhRvD..68h3002H} and the corresponding power spectrum \citep{2003PhRvD..67l3507K,2011PhRvD..83d3005H}. 
The non-Gaussian structure of the observed CMB due to lensing has been directly addressed by \citet{2000PhRvD..62d3007H}  and \citet{2011JCAP...03..018L} focusing on the CMB bispectrum, while \citet{2013MNRAS.428.2628M} employed skew-spectra and Minkowski Functionals. 
The reconstruction of the CMB deflection field is of great importance in removing lensing induced $B$-mode polarization in order to access primordial tensor perturbations \citep{2002PhRvL..89a1303K,2004PhRvD..69d3005S}. Furthermore, since the deflection field probes the integrated mass distribution between today's observer and the surface of last scattering it helps inventory the Universe's matter content \citep{2006PhRvD..73d5021L}.
 
All these applications of CMB lensing depend crucially on an accurate theoretical modeling of the lensing effect. Most work in this field concentrated on the exact computation of lensed power spectra \citep{1996ApJ...463....1S}, culminating in the all-sky expressions of \citet{2005PhRvD..71j3010C}. In addition to that, there are investigations of the corresponding two-point correlation function \citep[e.g.][]{1997ApJ...489....1M,2000ApJ...533L..83T}. The deflection field itself received comparatively little attention. \citet{2008MNRAS.388.1618C} constructed realistic CMB deflection maps from the Millennium Simulation. The resulting lensed CMB temperature and polarization maps agree very well with the theoretical prediction \citep{2009MNRAS.396..668C}. Using the results of \citet{2008MNRAS.388.1618C}, \citet{2011MNRAS.411.1067M} showed that the small amount of non-Gaussianity present in the CMB lensing excursion angle has only a marginal effect on the lensed CMB spectra. In a recent study \citet{2013JCAP...09..004C} produced lensing deflection maps from $N$-body simulations in the framework of interacting dark energy cosmologies. They found that for coupled dark energy models the (statistical) power of lensing distortions are either enhanced or suppressed (depending on the specific model under consideration) at the ten per cent level with respect to a $\Lambda$ cold dark matter ($\Lambda\mathrm{CDM}$) universe. 
Such differences are not necessarily restricted to the two-point function but may already be present at lower level. A particularly powerful one-point statistics is provided by the distribution of extreme values. 
Distinct properties of such a distribution then might be used to discriminate between different cosmologies. 
Furthermore, extreme value statistics are potentially more sensitive to deviations from Gaussianity than conventional higher-order statistics because the distribution of peaks in a Gaussian field declines rather fast. Non-Gaussianities, however, alleviate this sharp fall-off.
In this paper we investigate to which extent and under which conditions general extreme value statistics can be applied to the CMB deflection field in the context of standard $\Lambda\mathrm{CDM}$.

The outline of this paper is as follows: In Section~\ref{sec_methodology} we introduce the basic concepts of general extreme value statistics and explain its application to CMB lensing. Section~\ref{sec_results} compiles our results for both simulated Gaussian and reconstructed data. We conclude in Section~\ref{sec_conclusion}.

As reference we choose a $\Lambda\mathrm{CDM}$ cosmology with parameter values: $\Omega_\mathrm{CDM}\, h^2 = 0.1143$, $\Omega_\mathrm{b} \, h^2 = 0.02256$ and $\Omega_\Lambda = 0.721$ describing the Universe's energy budget; $A_\mathrm{s} = 2.457 \times 10^{-9}$ and  $n_\mathrm{s} = 0.96$ characterizing the primordial perturbations;  $h=0.701$ for today's Hubble constant, i.e. $H_0 = 100\, h \, \mathrm{km}\, \mathrm{s}^{-1}$.

\section{Methodology}
\label{sec_methodology}

\subsection{CMB lensing}

Due to Liouville's theorem gravitational lensing conserves the surface brightness of a given source \citep[e.g.][]{2001PhR...340..291B}. Thus, the lensed temperature contrast~$\tilde\Theta$ results from the primordial one via the remapping 
\begin{equation}
 \tilde\Theta ( \bmath{\hat n} )= \Theta [ \bmath{\hat n} + \bmath{d} (\bmath{\hat n}) ]
 \label{eq_remapping}
\end{equation}
\citep[see][for a comprehensive review on CMB lensing]{2006PhR...429....1L}.
The lensing deflection angle~$\bmath{d}$ is the angular gradient of the lensing potential
\begin{equation}
 \bmath{d} (\bmath{\hat{n}})= \nabla_{\theta,\phi} \psi ( \bmath{\hat{n}}) = 2 \, \nabla_{\theta, \phi} \int_0^{\chi^\star} \dd\chi \frac{\chi-\chi^\star}{\chi\chi^\star} \Psi (\bmath{\hat{n}}, \chi),
\end{equation}
where the latter follows from the line-of-sight projection of the gravitational potential~$\Psi$. Assuming a spatially flat universe we can use angular diameter and comoving distances interchangeably (denoted by~$\chi$). We further take recombination to be instantaneous, i.e the last scattering surface at~$\chi^\star$ is infinitely thin.
As detailed in the appendix of \citet{2005PhRvD..71h3008L} the remapping of the primordial CMB on the full sky involves the two components of the deflection angle~$d_{\theta,\phi}$ and its modulus~$d$ as well. In what follows we thus consider four different fields, both components of the spin-1 field~$\bmath{d}$ and the scalar fields~$\psi$ and~$d$.

\subsection{CMB lensing reconstruction}
\label{sec:reconstruction}

\begin{figure}
\centering
\resizebox{\hsize}{!}{\includegraphics{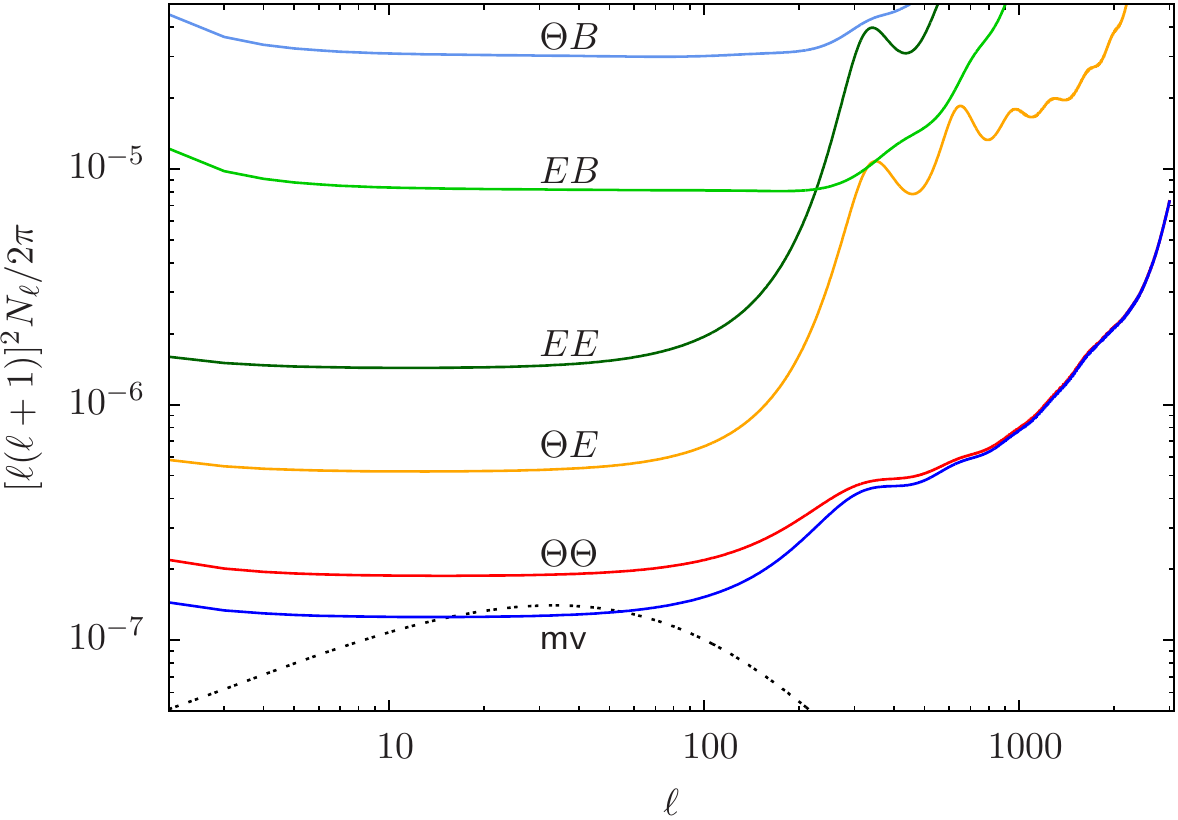}}
 \caption{Reconstruction noise of the quadratic estimator of the CMB lensing potential assuming a \textit{Planck}-like experiment \citep[cf.][]{2002ApJ...574..566H}. The noise levels correspond to different combinations of the CMB temperature ($\Theta$) and polarization ($E/B$-modes). The~$\Theta\Theta$-estimator is close to the minimum variance combination (mv) of all estimators due to the limited polarization sensitivity of \textit{Planck}. The black dashed curve indicates the power spectrum of the deflection field.}
 \label{fig:reconstruction_noise}
\end{figure}

The nonlinear nature of CMB lensing becomes apparent from equation~\eqref{eq_remapping}. The statistics of the lensed temperature fluctuations is therefore inevitably non-Gaussian and its covariance matrix acquires off-diagonal elements which are, to lowest order, linear in the lensing potential. Thus, suitably weighted they can be used to construct a quadratic estimator of the lensing potential \citep{2003PhRvD..67h3002O}. 
 This estimator is unbiased and optimal in the sense that its (Gaussian) variance is minimal. However, as  
 Figure~\ref{fig:reconstruction_noise} demonstrates for CMB experiments with sensitivity comparable to the \textit{Planck} satellite the reconstruction is noise dominated. Furthermore, we see that the estimator based on the temperature field alone performs best. The gain from including polarization data is negligible. We therefore focus on the $\Theta\Theta$-estimator and leave other combinations aside.
Once the lensing potential is reconstructed the components of the deflection angle are readily obtained by differentiation.

\subsection{General extreme value distribution}
Under rather general conditions the extrema of a set of random variates follow a distinct family of distribution functions \citep[e.g.][]{UBHD-65231677, 2004GEV}.
To be specific, consider the maximum~$M_n \equiv \mathrm{max} \left( X_1, \ldots, X_n\right)$ of a sequence of independent and identically distributed (i.i.d.) random variables~$X_i$. An analog to the central limit theorem states that in the limit~$n \rightarrow \infty$ the cumulative distribution function (CDF) of~$M_n$ approaches the general extreme value distribution (GEV) given by
\begin{equation}
	G_{\alpha,\beta,\gamma}(x) = \exp \left[ -\left( 1 + \gamma\, \frac{x - \alpha}{\beta} \right)^{-1/\gamma} \right]
	\label{eq:definition_gev}
\end{equation}
with location, scale and shape parameters~$\alpha$, $\beta$ and $\gamma$, respectively. The sign of the shape parameter determines the type of the GEV: Gumbel-type ($\gamma = 0$), Fr\'echet-type ($\gamma >0$) and Weibull-type ($\gamma < 0$). Analogous considerations hold true for the CDF of minima.
The corresponding probability density function (PDF) results from differentiating~equation~\eqref{eq:definition_gev}
\begin{equation}
 g_{\alpha,\beta,\gamma} ( x ) = \frac{\dd G_{\alpha,\beta,\gamma}(x)}{\dd x}.
\end{equation}
There are numerous applications of extreme value statistics in cosmology \citep[e.g.][considering strong and weak lensing, respectively]{2012A&A...547A..67W,2013arXiv1305.1485C}. A CMB related study has been carried out by \citet{2009MNRAS.400..898M}, while \citet{2011MNRAS.414.2436C} presented a general investigation of smooth Gaussian random fields in two and three dimensions.

\subsection{Sampling strategy}
\label{subsec:sampling_strategy}

\begin{figure}
\centering
\resizebox{\hsize}{!}{\includegraphics{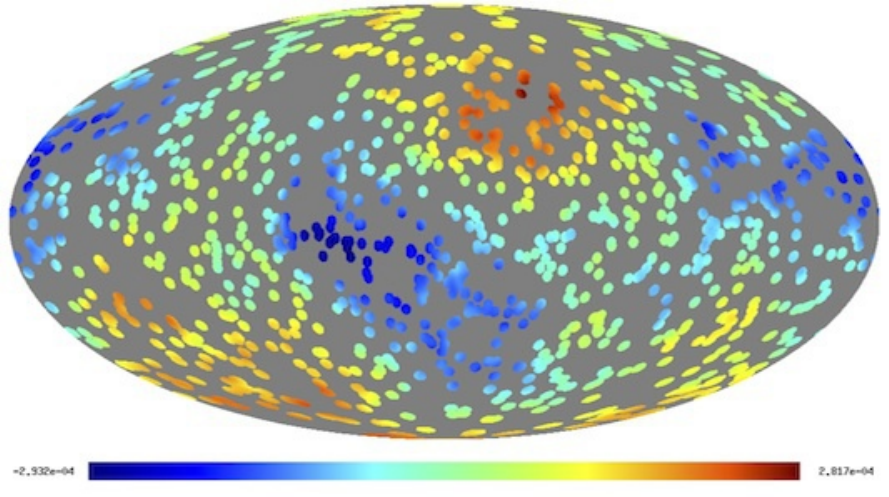}}
 \caption{Illustration of our sampling strategy. A map of the CMB lensing potential ist covered by 1000 patches. The radius of each patch is $2^\circ$.}
 \label{fig:sampling_strategy}
\end{figure}

In case of a collection of normal Gaussian variates the distributions of maxima and minima each approach the standard Gumbel-type CDF, i.e.~$G_{0,1,0}$. Both assumptions, Gaussianity and statistical independence, hold neither for the lensing potential nor the corresponding deflection field. The deviations from Gaussianity are small but the deflection field is correlated over the degree scale corresponding to the angular size of a typical lens which is located half the way to the last scattering surface \citep[cf.][]{2006PhR...429....1L}. These correlations can be partially removed by applying a low-pass filter to the harmonic representation of the lensing potential
\begin{equation}
 \psi_{\ell m} = \int \dd \Omega \, Y^*_{\ell m} (\bmath{\hat{n}})\, \psi(\bmath{\hat{n}}).
\end{equation}
Dividing the smoothed maps into patches we can estimate the PDF of extreme values by determining the maximum, respectively minimum, in each patch. A faithful estimate put two major constraints on the choice of patches. On the one hand they have to be extended enough to contain a sufficiently large number of deflection angles in order to make the sample converge to the limiting CDF. On the other hand the patches must not overlap but have to be as numerous as possible to sample the PDF appropriately. In order to satisfy both requirements, we cover the smoothed map homogeneously with 1000 discs the radius of which we choose to be $2^\circ$. This corresponds to a $\sim$30 per cent coverage of the whole sky. The correlation coefficient between two patches has been estimated numerically from a sample of reconstructions and is typically~$\sim$0.3. An exemplary collection of patches is shown in Figure~\ref{fig:sampling_strategy}. The coordinates of the discs' centers are drawn from a random distribution being uniform over~$\mathbb{S}^2$.

\section{Results}
\label{sec_results}

\subsection{Gaussian data}
\label{sub_sec:Gaussian_data}
We start our analysis with a sample of 100 Gaussian realizations of the lensing potential and the corresponding deflection field.
For the generation of the synthetic data we first compute the lensing potential power spectrum~$C_\ell^{\psi\psi}$ using \textsc{camb}\footnote{\url{http://camb.info/}} \citep{2000ApJ...538..473L}. Then we invoke the \textsc{healpix}\footnote{\url{http://healpix.sourceforge.net/}} package \citep{2005ApJ...622..759G} to generate a set of harmonic coefficients such that~$\left\langle\psi_{\ell m} \psi^*_{\ell' m'} \right\rangle = C_\ell^{\psi\psi}\delta_{\ell\ell'}\delta_{mm'}$. Subsequently, these coefficients are used to build full-sky maps of the lensing potential and its first (angular) derivatives with \textsc{healpix} resolution parameter $N_\mathrm{side}=2048$, which captures angular scales of~$\sim$1.72\arcmin. During the map making procedure we include multipoles up to~$\ell_\mathrm{max}^{\mathrm{sim}}=3000$. 

In the analysis we choose a sharp cut-off in the harmonic domain $(\ell_\mathrm{max}=100)$ as filter function. Following the sampling procedure outlined above we estimate the distribution function of the extrema for each realization. We fit the GEV~\eqref{eq:definition_gev} to the data allowing all three parameters to vary. Shape, location and scale parameters are obtained by maximizing a log-likelihood function. A typical fit result is illustrated in 
Figure~\ref{fig:histogramm_reconstruction} for the maxima of the $\theta$-component of the lensing deflection angle. Note that data (grey boxes) and fit (red dashed curve) in this figure are derived from a realistic reconstruction (which we discuss at length in Section~\ref{subsec:reconstructed_data}) instead of a Gaussian realization. Maxima are generally denoted by a superscripted '+', e.g.~$\psi^+$ for the lensing potential, while we use a superscripted '$-$' in case of minima.

\begin{figure}
\centering
\resizebox{\hsize}{!}{\includegraphics{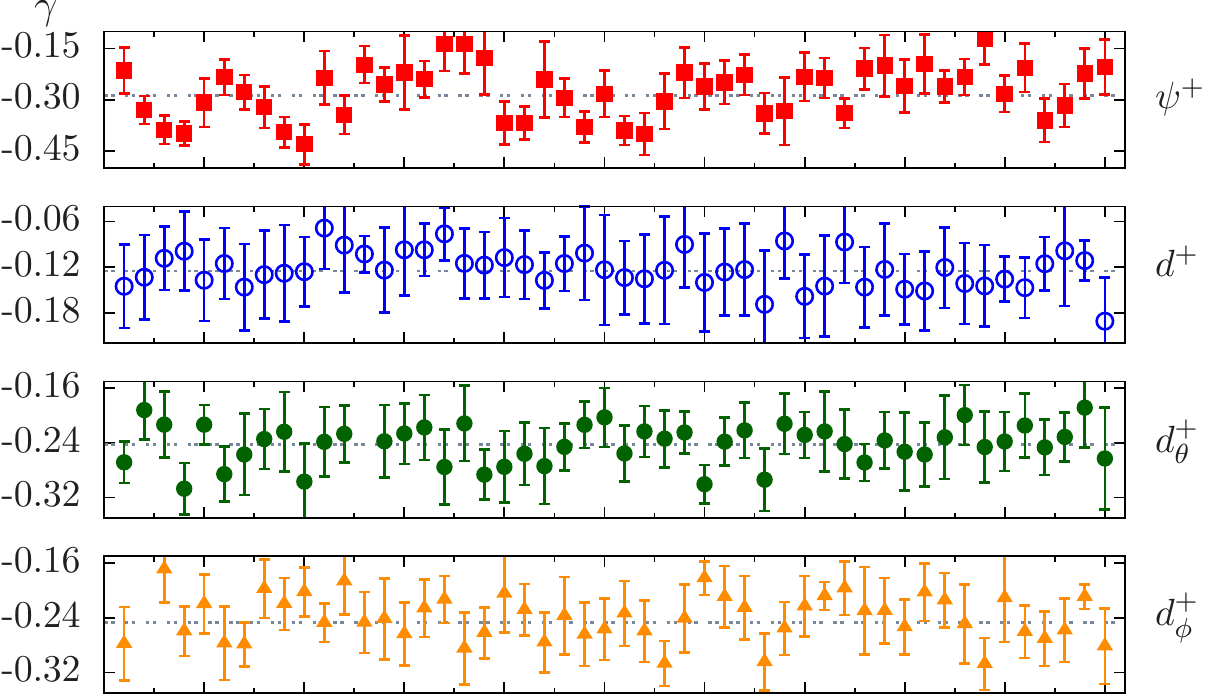}}

 \vspace{0.175cm}
 \resizebox{\hsize}{!}{\includegraphics{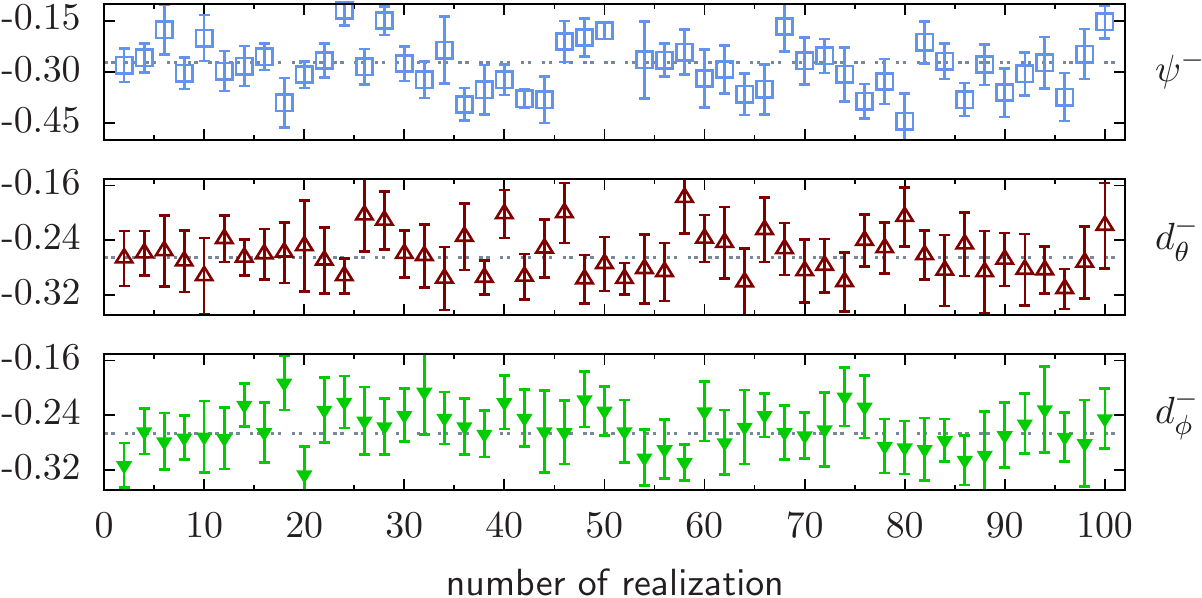}}
 \caption{Fit results for the shape parameter~$\gamma$ of the GEV considering the maxima ('+') and minima ('$-$') of the lensing potential~($\psi^\pm$), the modulus~($d^+$) and components~($d_{\theta,\phi}^\pm$) of the corresponding deflection field. The minima of the modulus do not follow a GEV function. The error bars correspond to the estimated $1\sigma$ error of the fit, while the grey dashed line indicates the mean value. We only show half of the realizations for clarity.}
 \label{fig:fit_gamma_maxima_and_minima}
\end{figure}

We investigate the fit results in a more quantitative way in Figure~\ref{fig:fit_gamma_maxima_and_minima}. Here we concentrate on the shape parameter. We find that for all four observables the distribution of maxima is consistently described by the Weibull-type GEV. The (relative) scatter, however, is large. It is largest in case of the lensing potential and the deflection angle modulus, where it amounts to about 50 per cent. This compares to approximately 30 per cent in case of the deflection angle components. As expected from statistical isotropy the shape parameter is very much the same for the two deflection angle components $(\gamma \sim -0.24)$ taking the fit uncertainty into account. The shape parameter of the modulus, which is bound by zero from below, is markedly larger, i.e. less negative $(\gamma\sim -0.12)$, whereas we find a smaller value $(\gamma \sim -0.29)$ for the lensing potential. Recovering a GEV of Weibull form agrees with the general considerations of \citet{2011MNRAS.414.2436C}.

Concerning the distribution of minima we find comparable results. They are well described by the Weibull-type (see Figure~\ref{fig:fit_gamma_maxima_and_minima}) and the scatter is of the same order. As expected, the minima of the modulus do not follow the GEV due to its lower bound. The poor fit results of the shape parameter are omitted.

\begin{figure}
 \centering
 \resizebox{\hsize}{!}{\includegraphics{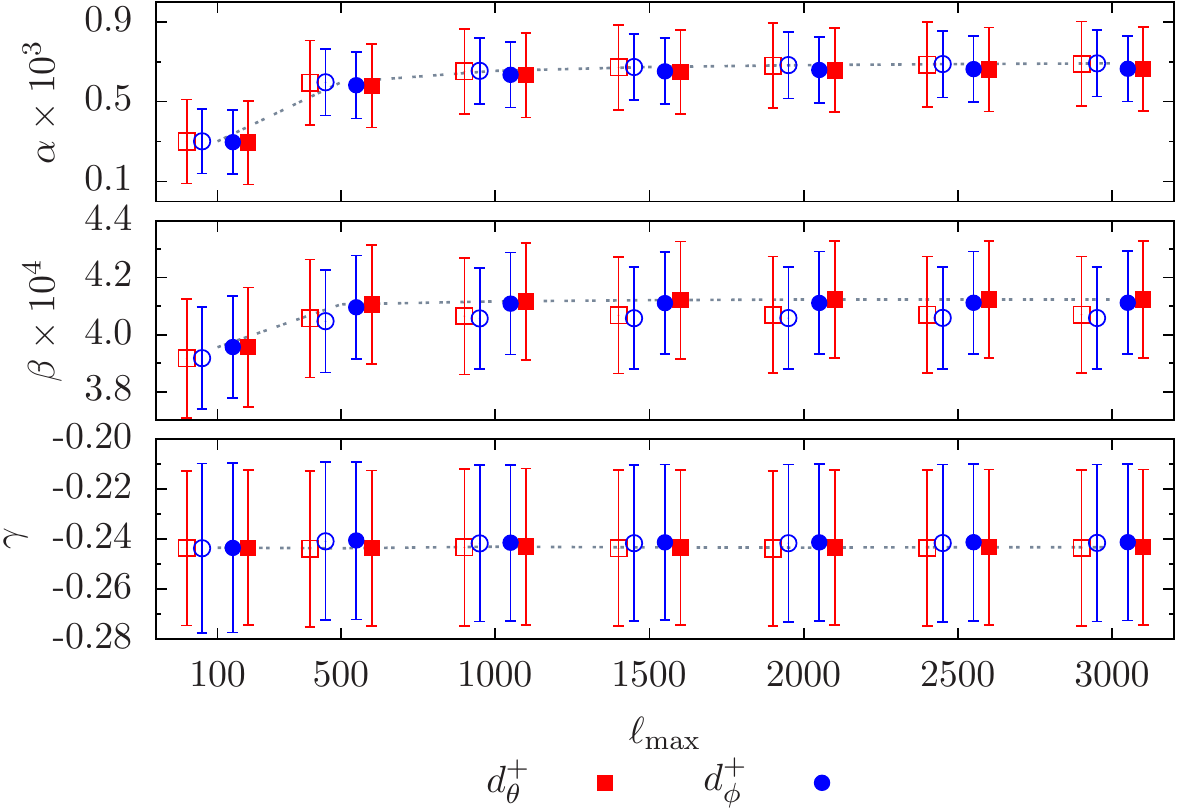}}
 \caption{Location~$\alpha$, scale~$\beta$ and shape parameter~$\gamma$ (from top to bottom) for the distribution of maxima of the deflection angle components as a function of the cut-off in the harmonic domain. The error bars correspond to the $1\sigma$ error over 100 Gaussian realizations and have been artificially amplified by a factor of ten in case of the location parameter. Filled symbols indicate our reference model, while light ones denote the \textit{Planck} best-fitting cosmology.}
 \label{fig:fit_alpha_beta_gamma_maxima_different_lmax}
\end{figure}

\begin{figure}
 \resizebox{\hsize}{!}{\includegraphics{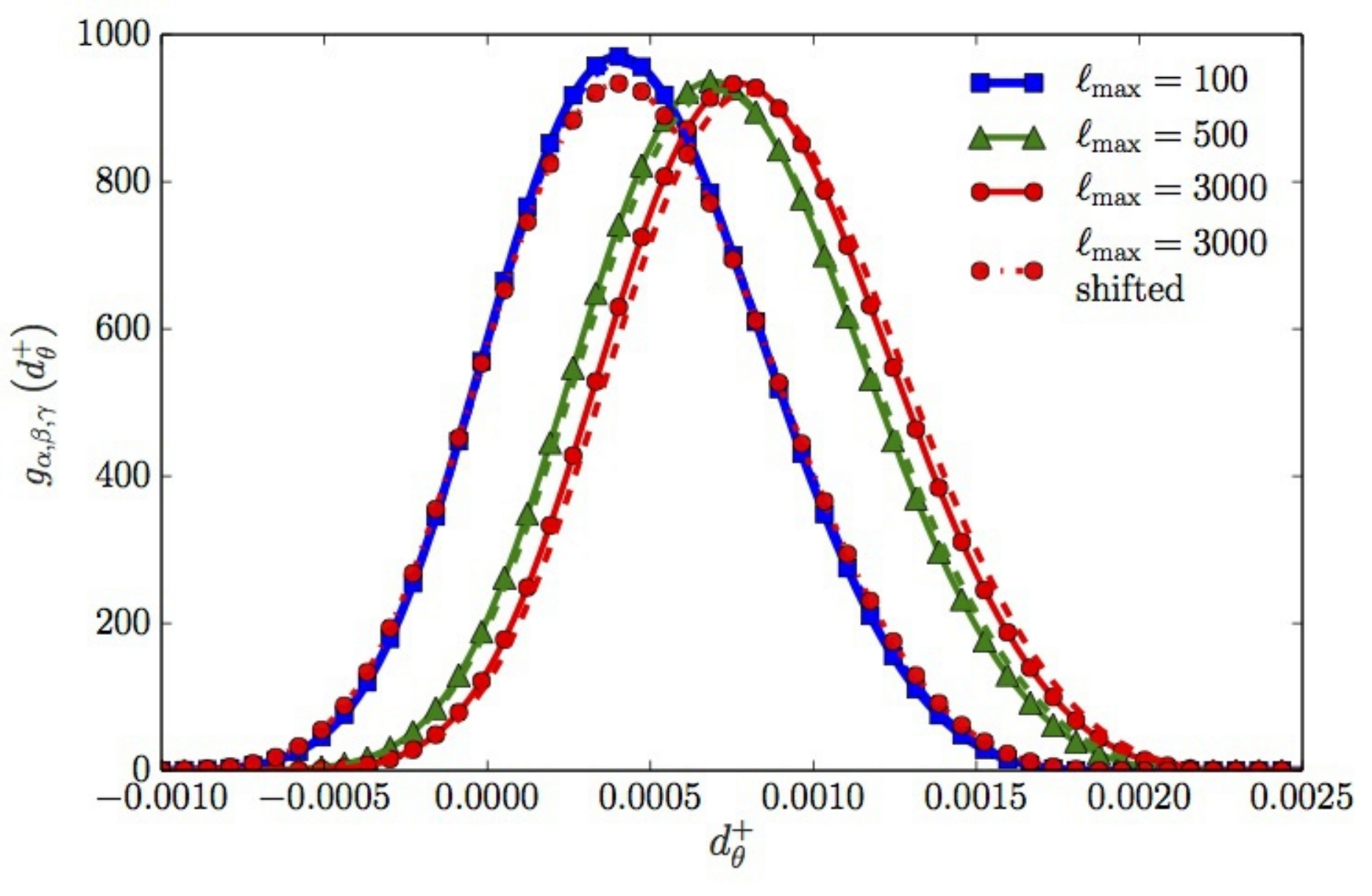}}
 \caption{Distribution functions of deflection angle maxima for different cut-off choices. The shifted curve for~$\ell_\mathrm{max}=3000$ has the same location parameter as that found for~$\ell_{\mathrm{max}}=100$ and illustrates the broadening of the distribution with increasing cut-off size. Dashed lines indicate the \textit{Planck} best-fitting cosmology.}
  \label{fig:distributions_plot}
\end{figure}

In a next step we consider the influence of the cut-off choice. We reexamine the GEV of our sample for different filter scales. We aim at rather general properties which are independent of a particular realization. Recalling the spread in shape parameters we believe that the deflection field components are most likely to comply with the requirement of generality. We therefore leave the other two observables aside in the following. 
In addition to the degree-size cut-off we investigate the range~$\ell_\mathrm{max} = 500, \ldots, 3000$  in steps of 500. 
For each cutoff we plot the mean over the entire sample in Figure~\ref{fig:fit_alpha_beta_gamma_maxima_different_lmax}. 
The corresponding distributions functions are shown in Figure~\ref{fig:distributions_plot}.
The error bars in Figure~\ref{fig:fit_alpha_beta_gamma_maxima_different_lmax} indicate the standard deviation which is at the 5-15 per cent level. Smallest (largest) scatter is observed in the location (shape) parameter (note that for the location parameter the errors are artificially amplified by a factor of ten in the plot).  Figure~\ref{fig:fit_alpha_beta_gamma_maxima_different_lmax} reveals that the results for the shape parameter are rather insensitive to the cut-off choice. The other parameters of the GEV depend much more strongly on the filter scale. In particular we observe that removing small-scale fluctuations results in smaller location and scale parameters. Thus, mean and width of the corresponding GEV decrease (cf. Figure~\ref{fig:distributions_plot}). This behaviour is expected because the degree-size cut-off has been introduced to weaken the correlations in the lensing field. Its application removes inevitably the contributions of the rather efficient degree-sized lenses.
Finally, we would like to emphasize that our findings presented so far do not show any significant differences for both our reference model and the \textit{Planck} best fitting cosmology \citep{2014A&A...571A..16P}. 
The main difference between these two parameter sets, which is particularly relevant for lensing, is that in case of \textit{Planck} the matter density is enhanced, whereas  the amplitude of the primordial fluctuations is reduced.
Considering the minima leads to the same conclusions.

\begin{figure}
 \resizebox{\hsize}{!}{\includegraphics{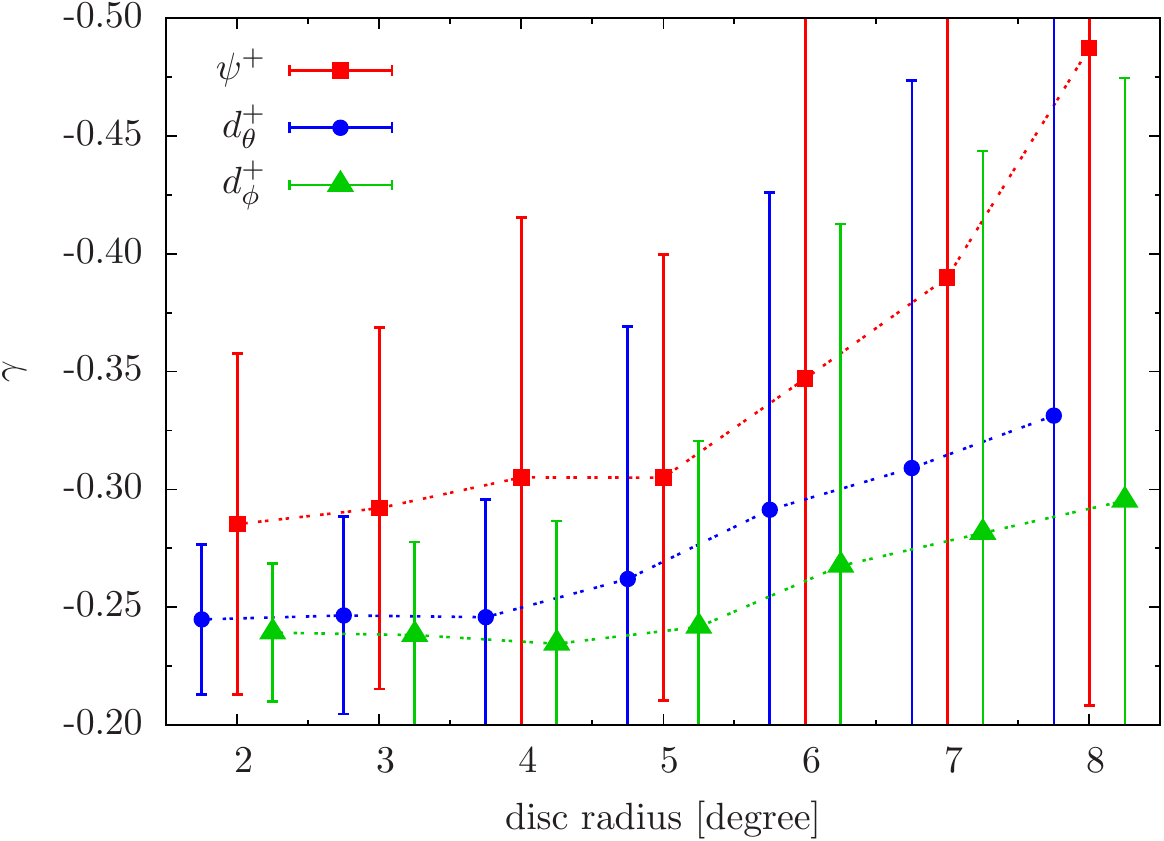}}
 \caption{Fit results for the GEV shape parameter~$\gamma$ of the maxima of the CMB lensing potential and both deflection field components as a function of disc radius, i.e. patch size, used for sampling. Error bars indicate $1\sigma$~errors. Note the artificial offset in radius we included for clarity in case of the deflection field components.}
 \label{fig:shape_parameter_different_patch_sizes}
\end{figure}

We next demonstrate that our results are independent of the patch size chosen for the sampling process explained in Section~\ref{subsec:sampling_strategy}. In Figure~\ref{fig:shape_parameter_different_patch_sizes} we show the fitting results for the shape parameter as a function of patch size. As before we use 100 maps of the deflection field and lensing potential, respectively, which are smoothed in harmonic space at~$\ell_{\mathrm{max}} = 100$. The disc radii range from $2^\circ$ to $8^\circ$, while the percentage of coverage is kept fixed to be 30 per cent. Thus, the larger the radius the smaller the number of discs covering the map in order to avoid overlapping patches. From Figure~\ref{fig:shape_parameter_different_patch_sizes} it can be seen that the shape parameter hardly varies once the sampling process is able to resolve the limiting distribution.

\subsection{Reconstructed data}
\label{subsec:reconstructed_data}
Having found that the distribution of extrema in Gaussian realizations of the CMB lensing deflection field is well fitted by a Weibull GEV we now leave these idealized conditions behind and address the more realistic case of reconstructed deflection fields. To this end we generate 100 lensed CMB maps using \textsc{lenspix}\footnote{\url{http://cosmologist.info/lenspix/}} \citet{2005PhRvD..71h3008L} and subsequently apply the optimal quadratic estimator described in Section~\ref{sec:reconstruction} to recover the lensing potential. In the framework of \textsc{lenspix} lensed maps are generated by remapping the pixels of unlensed CMB maps in combination with a bicubic interpolation scheme. We choose the resolution of the unlensed maps  to be a factor of 1.5 higher than that of the lensed maps for which we use \textsc{healpix} resolution~$N_\mathrm{side}=2048$ and set~$\ell_\mathrm{max}^\mathrm{sim}=3000$. The initial CMB power spectrum from which we synthesize the input maps is computed using again \textsc{camb}. 
Note that \textsc{lenspix} uses a purely Gaussian realization of the lensing potential to construct the lensed CMB maps because it only considers the power spectrum of the potential; no higher statistics are included. Hence, the lensing potential realizations considered here differ from those of Section~\ref{sub_sec:Gaussian_data} by the noise due to the estimator (cf.~Figure~\ref{fig:reconstruction_noise}).

\begin{figure}
 \centering
 \resizebox{\hsize}{!}{\includegraphics{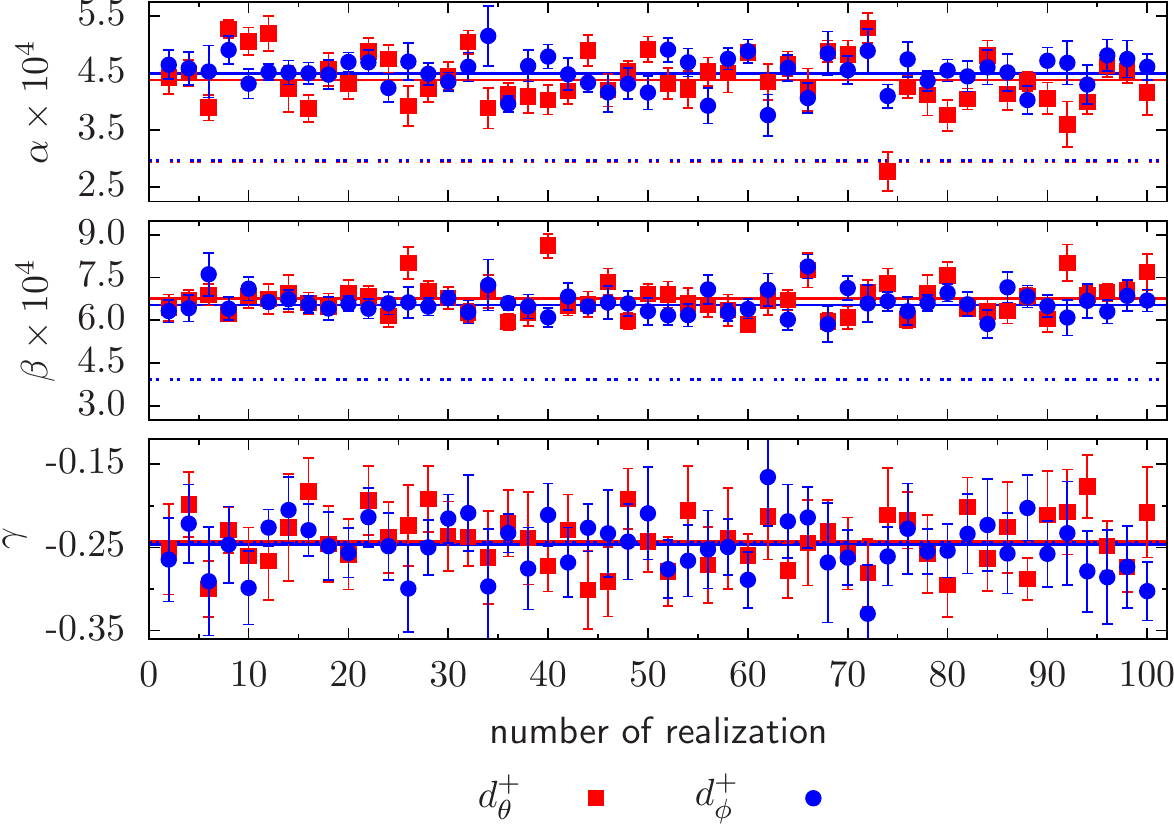}}
 \caption{Fit results for the location, scale and shape parameters of the GEV considering the maxima of 100 reconstructed deflection angle maps. The error bars indicate an estimate of the fit uncertainty. Solid lines represent the average over all realizations, while the dashed ones correspond to the mean values found from Gaussian realizations.}
  \label{fig:fit_results_alpha_beta_gamma_reconstructions}
\end{figure}

\begin{figure*}
\includegraphics[scale=0.35]{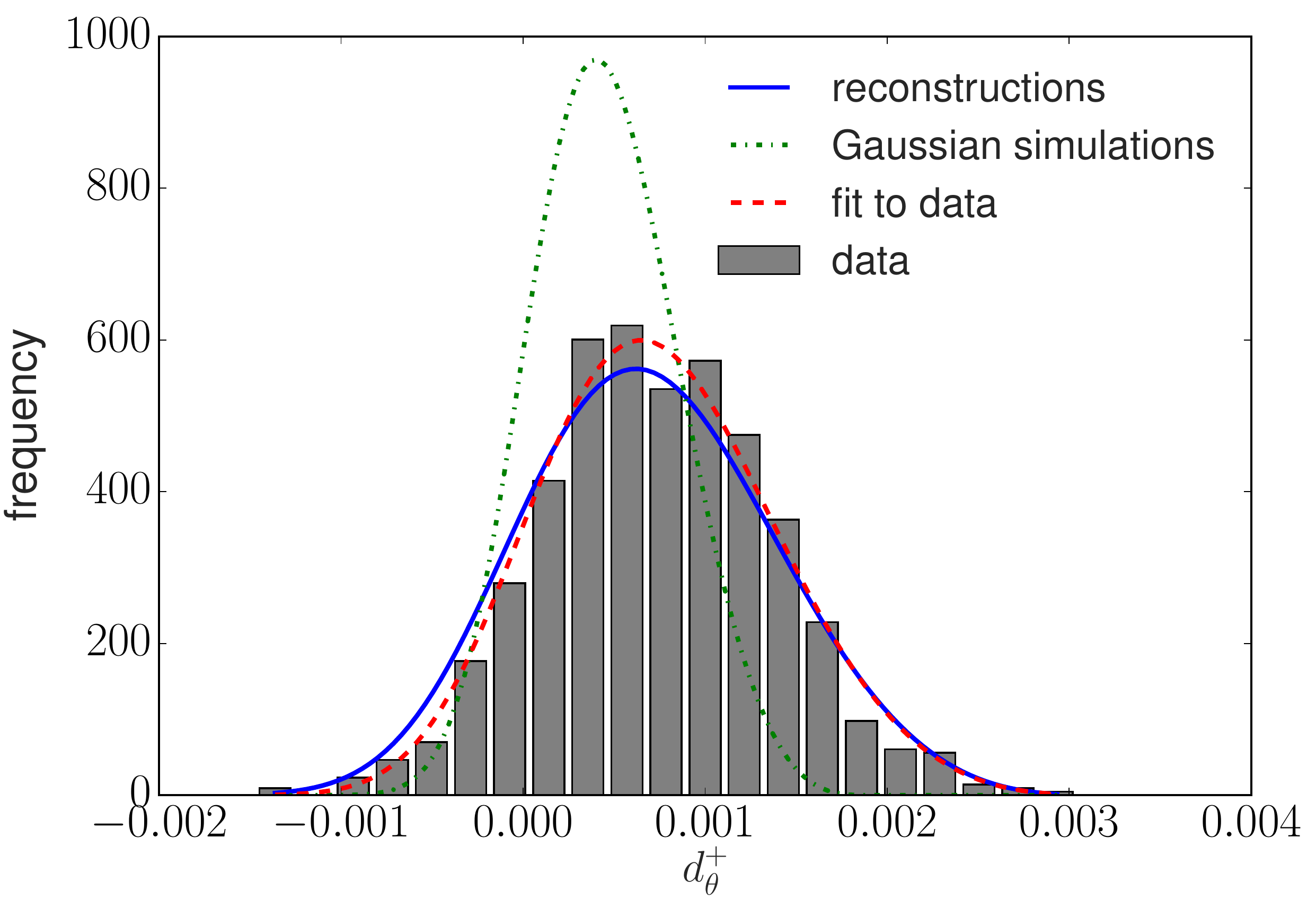}\hfill
\includegraphics[scale=0.35]{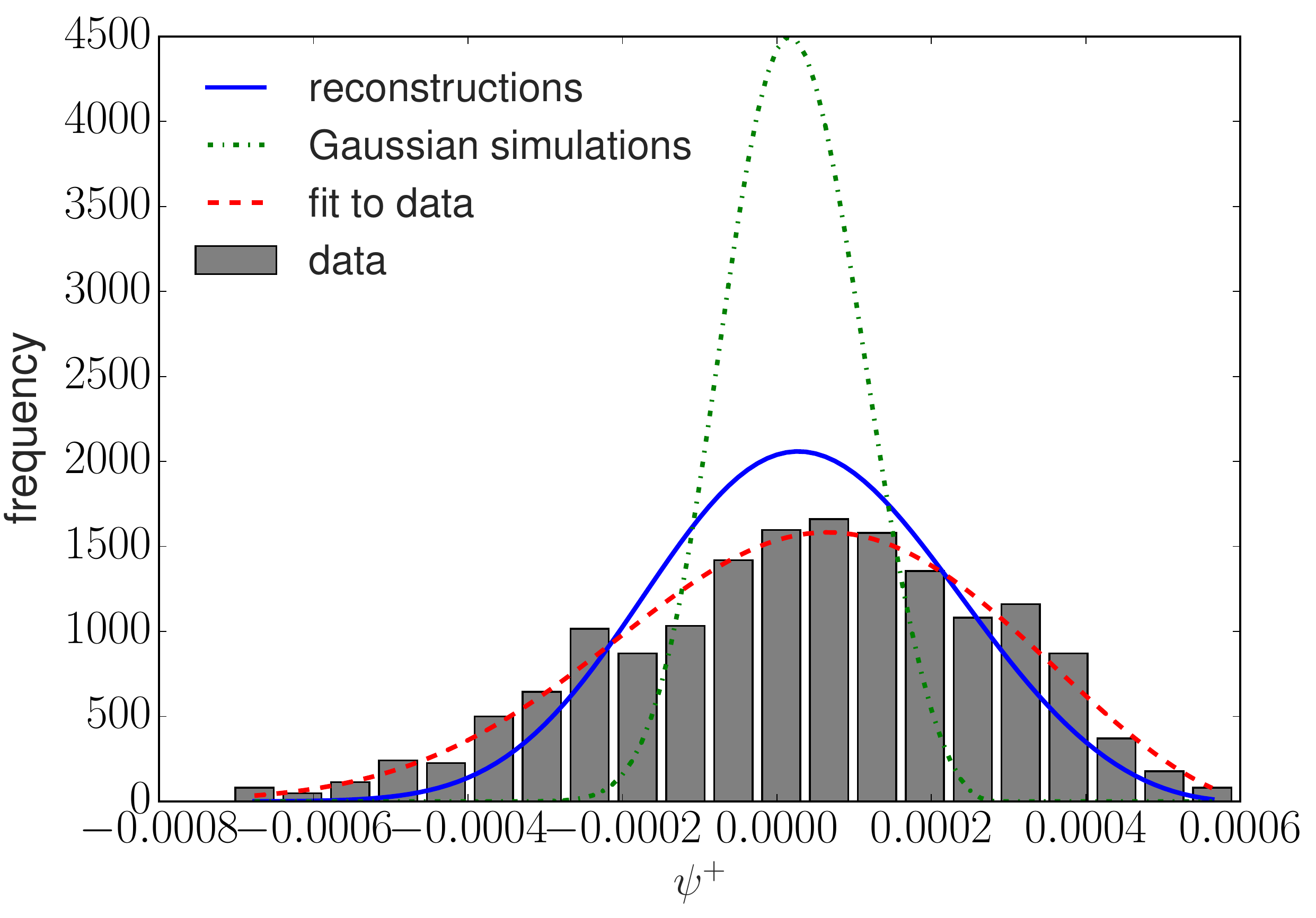}
  \caption{Illustration of the PDF of maxima of the $\theta$-component of the reconstructed CMB deflection angle (left panel) and the corresponding lensing potential (right panel). The data is well fitted by a Weibull GEV when its parameters are inferred from the data itself (dashed curve). The disagreement is apparent when the averaged shape, location and scale parameters obtained from either 100 Gaussian realizations (dashed dotted curve) or  100 reconstructions (solid curve) are used in stead.}
\label{fig:histogramm_reconstruction}
\end{figure*}

This noise makes a low-pass filter increasingly important. Since the deflection field is derived by differentiation from the lensing potential, there is a simple algebraic relation between the harmonic coefficients of the potential and those of the spin-1 components of the deflection field \citep[e.g.][]{2005PhRvD..71h3008L}: ${}_1\bmath d_{\ell m} = \sqrt{\ell(\ell +1 )}\, \psi_{\ell m}$. Thus, including the quite noisy small scale modes in the potential reconstruction introduces a substantial bias because the noise leads to a spurious enhancement of the deflections and thus widens the GEV. This enhances artificially both location and scale parameter. They increase by a factor of about 1.5 - 2. The shape parameter is only marginally affected, the distribution of maxima follows a Weibull-type GEV. We illustrate these findings in Figures~\ref{fig:fit_results_alpha_beta_gamma_reconstructions} and~\ref{fig:histogramm_reconstruction} (left panel). The scatter observed in Figure~\ref{fig:fit_results_alpha_beta_gamma_reconstructions} is compatible to the case of Gaussian realizations (cf. Figure~\ref{fig:fit_gamma_maxima_and_minima}). 
The right panel of Figure~\ref{fig:histogramm_reconstruction} reveals that in case of the reconstructed lensing potential the discrepancy with respect to the mock data is even larger. The GEV of (both) potential maxima and minima obtained from Gaussian simulations is (are)  much narrower than that(those) for the reconstructed data.

\begin{figure}
 \centering
 \resizebox{\hsize}{!}{\includegraphics{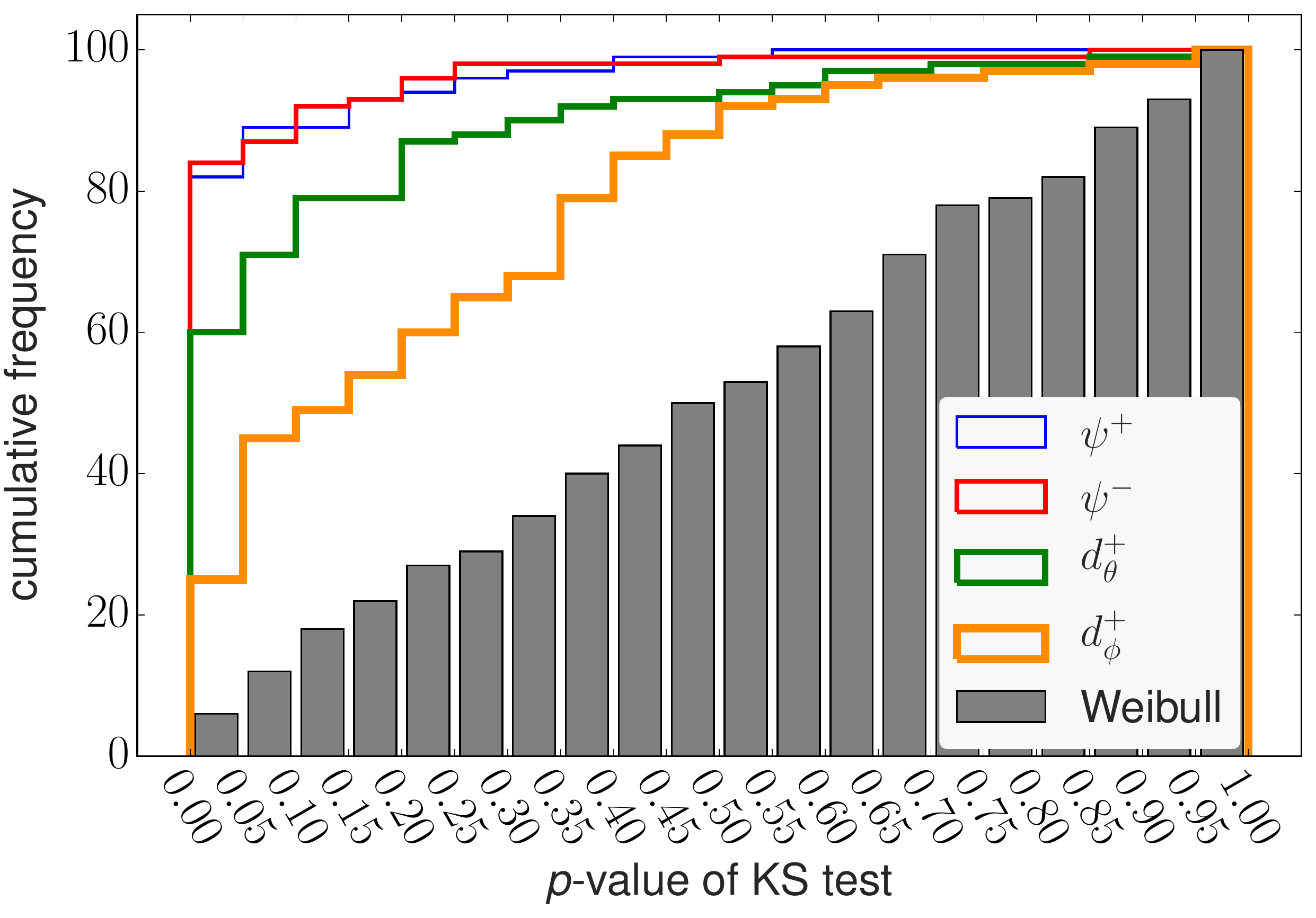}}
 \caption{Cumulative frequency of the outcome of Kolmogorov-Smirnov (KS) test results. The distribution of maxima in the reconstructed CMB deflection angle components has been tested against a Weibull GEV the parameters of which have been inferred from a different data set (thick curves). In addition we show the outcome for both maxima and minima in the reconstructed lensing potential (light curves). The grey boxes indicate the outcome for samples drawn from the assumed GEV.}
 \label{fig:ks_test_results}
\end{figure}

These findings make, to say the least, a unified description of the maxima distribution in terms of the averaged shape, location and scale parameters rather questionable. In order to address our concerns quantitatively we investigate the goodness of fit employing a Kolmogorov-Smirnov (KS) test. It is a non-parametric test and in order to get meaningful results the distribution parameters must not be estimated from the same data set the test is applied to. We therefore generate a second sample of 100 reconstructions which are completely independent of the one presented before and from which the distribution parameters were determined. Both samples have only the background cosmology in common. The outcome of the test, i.e. the KS supremum values between proposed and fitted CDF, can be translated into a more readily accessible \textit{p}-value. Usually, a \textit{p}-value less than five per cent is regarded as an indicator that there is strong presumption against the null hypothesis, that the data follows the proposed distribution (in our case a Weibull GEV parametrized by the average shape, location and scale parameters), whereas an outcome greater than ten per cent implies no presumption against the null hypothesis. We compile the test results in Figure~\ref{fig:ks_test_results}. For comparison we also show the results for 100 data sets which have been sampled from the assumed Weibull distribution. Considering the $\theta$-component only $\sim40$ ($\sim25$) per cent of the reconstructed maps exceed the threshold of 0.05 (0.1). The agreement is better for the other component, where $\sim 75$ ($\sim 55$) per cent of the fit results pass the first (second) significance level. However, this needs to be compared to the outcome of the mock data. Not more than ten per cent of it fall below a significance level of 0.1.
Thus, the distribution of maxima of the CMB deflection angle components is consistently described by a Weibull-type GEV. But we find no evidence for a universal set of shape, location and scale parameters independent of the particular realization.

In order to investigate whether the lensing potential suffers from the same lack of universality we repeat the KS test for the extrema of the reconstructed potential. The results are shown in Figure~\ref{fig:ks_test_results}, too. Only a tiny fraction of fits passes the test; roughly 20 (10) per cent of the fit results are above the first (second) significance level. This is in line with our previous result that the (relative) scatter in the shape parameter is much larger for the lensing potential than for the deflection field (cf. Section~\ref{sub_sec:Gaussian_data}).

Despite the missing universality we would finally like to roughly estimate under which conditions the GEV of CMB lensing angles might help find deviations from a $\Lambda\mathrm{CDM}$ cosmology.  A shape parameter which exceeds clearly the typical fluctuations seen in Figure~\ref{fig:fit_results_alpha_beta_gamma_reconstructions}, say $\gamma\sim-0.1$ or~$-0.4$ to be specific, would hint at a different cosmological model. Such a model, however, must have rather distinct lensing properties (compared to $\Lambda\mathrm{CDM}$): Assuming that both location and scale parameters are unchanged the expectation value of the maximum deflection angle would become ten per cent larger or smaller, respectively. The variation is less pronounced  when the scale parameter, too, changes at a comparable level. For the detection of such strong variations in the lensing efficiency, however, cosmic shear measurements are much more appropriate.

\section{Conclusion}
\label{sec_conclusion}

We have applied general extreme value statistics to CMB lensing. We have found that both maxima and minima of the lensing potential and the deflection angle components derived from it follow a GEV of Weibull form. This is also true for the maxima of the deflection field modulus. A basic assumption of general extreme value statistics is statistical independence. We therefore employed a sharp cut-off in harmonic space in oder to suppress correlations in the lensing potential. Low-pass filtering becomes particularly important when a realistic but noisy reconstruction scheme for the lensing potential, which is not directly observable, is employed. Since shape, location and scale parameters of the GEV vary considerably between individual realizations of the deflection field the statistics' practical use as an additional cosmological probe is rather limited.

% --- Acknowledgements --- %
\section*{Acknowledgements}
We would like to warmly thank Sven Meyer for helpful discussions on the \textsc{healpix} package.
We are also grateful to the anonymous referee for her/his valuable comments.

\bibliography{bibtex/aamnem,bibtex/references}
\bibliographystyle{mn2e}

%\appendix

\bsp

\label{lastpage}

\end{document}